\documentstyle[12pt,epsfig]{article}
\def\simgt{\rlap{\lower 3.5 pt \hbox{$\mathchar \sim$}} \raise 1pt
  \hbox {$>$}}
\def\ppb{p\bar{p}}
\def\as{\alpha_s}
\def\gg{\tilde{g}\tilde{g}}
\def\ms{m_{\tilde q}}
\def\mg{m_{\tilde g}}
\def\md{m_{-}}
\catcode`@=11
\def\citer{\@ifnextchar [{\@tempswatrue\@citexr}{\@tempswafalse\@citexr[]}}

\def\@citexr[#1]#2{\if@filesw\immediate\write\@auxout{\string\citation{#2}}\fi
  \def\@citea{}\@cite{\@for\@citeb:=#2\do
    {\@citea\def\@citea{--\penalty\@m}\@ifundefined
       {b@\@citeb}{{\bf ?}\@warning
       {Citation `\@citeb' on page \thepage \space undefined}}%
\hbox{\csname b@\@citeb\endcsname}}}{#1}}
\catcode`@=12

\begin{document}
\thispagestyle{empty}

\hfill\vbox{\hbox{\bf DESY 95-104}
            \hbox{May 1995}
                                }
\vspace{1.in}
\begin{center}
\renewcommand{\thefootnote}{\fnsymbol{footnote}}
{\large\bf Gluino-Pair Production at the Tevatron}\\
\vspace{0.5in}
W.~Beenakker$^{1,2}$, R.~H\"opker$^1$,
M.~Spira$^3$ and P.~M.~Zerwas$^1$ \\
\vspace{0.5in}
$^1$ Deutsches Elektronen-Synchrotron DESY, D-22603 Hamburg, Germany \\
$^2$ Instituut-Lorentz, University of Leiden, The Netherlands
\footnote[3]{present address} \\
$^3$ II. Institut f.~Theor.~Physik \footnote[4]{Supported by
  Bundesministerium f\"ur Bildung und Forschung (BMBF), Bonn, Germany,
  under Contract 05 6 HH 93P (5) and by EU Program {\it Human Capital
    and Mobility} through Network {\it Physics at High Energy
    Colliders} under Contract CHRX--CT93--0357 (DG12 COMA).},
Universit\"at Hamburg, D-22761
Hamburg, Germany\\
\end{center}
\vspace{5cm}

\begin{center}
ABSTRACT \\
\end{center}
The next-to-leading order QCD corrections to the production of gluino
pairs at the Tevatron are presented in this paper. Similar to the
production of squark-antisquark pairs, the dependence of the cross
section on the renormalization/factorization scale is reduced
considerably by including the higher-order corrections. The cross
section increases with respect to the lowest-order calculation
which, in previous experimental analyses, had been evaluated at the
scale of the invariant energy of the partonic subprocesses.

\pagebreak

\section{Physical Set-Up}

The novel colored particles in supersymmetric theories, squarks and gluinos,
can be searched for most efficiently at high-energy hadron colliders,
i.~e.~the Tevatron $\ppb$ collider and the LHC $pp$ collider in the
future. The most
stringent lower bounds on squark and gluino masses have been set by
the Tevatron experiments CDF and D0. At the $90\%$ CL, the lower limit
of the gluino mass was found to be $157$ GeV, independent of the
squark mass \cite{massold,massnew}.

The cross sections for the production of squarks and gluinos in hadron
collisions were predicted at the Born level already quite some
time ago \citer{born1,isajet}. Only recently the
predictions have been improved by the
next-to-leading order QCD contributions for squark-antisquark pair
production \cite{been}. In a first step, the {\it gluon} radiative
corrections to
gluino pair production have been calculated in Ref.\cite{top}; these
corrections are closely related to the gluon corrections for
quark-antiquark production, requiring just the appropriate change of
the $SU(3)$ Casimir invariants. In this paper we shall present a
complete next-to-leading ${\cal O}(\as)$ analysis for gluino pair
production,
\begin{equation}
  \ppb \to \gg +X
\end{equation}
including the virtual effects of gluinos and squarks besides gluons
and quarks.

The technical set-up of the calculation is analogous to the squark
analysis. For the sake of simplicity all squark states are taken mass
degenerate. The only free parameters are therefore the masses of
squarks and gluinos, $\ms$ and $\mg$, respectively. (The top mass is
fixed to $176$ GeV \cite{topCDF,topD0}.)

The calculation has been performed in the Feynman gauge and the
singularities have been isolated by means of dimensional
regularization. The masses are renormalized in the on-shell
scheme. The massive particles are decoupled smoothly for momenta
smaller than their masses in the {\it modified} $\overline{MS}$ scheme
\cite{msbarext}. The soft and the hard gluon radiation are separated
by a small invariant mass cut-off
parameter $\Delta$ for the gluino-gluon and gluino-light quark final
states, regularizing the infrared divergences after the virtual
corrections are included. If soft and hard gluon emissions are added
up, the $\Delta$ dependence disappears from the total cross section
for $\Delta \to 0$. The remaining collinear mass singularities are
absorbed in the renormalization of the parton densities carried out in the
$\overline{MS}$ scheme \cite{altar}. We have adopted the GRV \cite{GRV} and
CTEQ(2pM) \cite{CTEQ} parametrizations of the parton densities to compare
the NLO predictions with the LO approximations; the MRS(H) \cite{MRS}
parametrizations have been included finally to assess the
uncertainties of the predictions associated with different
parametrizations of the parton densities.

If the gluinos are lighter than the squarks, on-shell squarks can
decay into quarks plus gluinos, $\tilde{q} \to q\tilde{g}$. We will
assume in the present analysis that these events are attributed to
squark final states; technically, we subtract off the squark-gluino
production cross section $qg \to \tilde{q}\tilde{g}$ after including
the non-zero squark width in the propagator. Analogously, the
decay of gluinos to squarks is disregarded in the wedge $\mg >
\ms$. Spurious singularities in higher orders at the decay threshold for
gluino decays to stop and top, can be regularized by taking the
non-zero widths of the particles into account and by performing the mass
renormalization of the gluino at the complex pole of the propagator.

The dominant contributions to the production cross sections of gluino
pairs in $\ppb$ collisions are due to the $gg \to \gg$ and the
$q\bar{q} \to \gg$ subprocesses. In contrast to squark-pair
production, the $gg$ initial state is dominant if both $\mg$ and $\ms$
are less than about 200 GeV. The $qg$ initial states give rise to
$\gg$ pairs only at next-to-leading order while the contribution of
quark-quark pair initial states to gluino pairs are of yet higher order.

\subsection{Gluon-gluon initial states}

To lowest order the diagrams contributing to the subprocess $gg\to\gg$
are analogous to the diagrams for quark-pair production, Fig.~1a. Typical
standard QCD and supersymmetric vertex corrections are
displayed in Fig.~1c. Gluon radiation processes, Fig.~1d, are added
incoherently.

Evaluating these diagrams analytically we obtain the double
differential cross sections $d\hat{\sigma}_{ij}/d\hat{t}d\hat{u}$ at the
parton level; $i,j$ are the parton indices $g,q,\bar{q}$ and
$\hat{t},\hat{u}$ the Mandelstam momentum transfer variables. The
total cross sections may be expressed in terms of scaling functions
$f_{ij}$,
\begin{equation}
  \hat{\sigma}_{ij} = \frac{\as^2(Q^2)}{\mg^2}\left\{f^{(0)}_{ij} + 4 \pi
  \as(Q^2) \left[f^{(1)}_{ij} + \bar{f}^{(1)}_{ij} \log \left(\frac{Q^2}{\mg^2}
\right) \right] \right\}
\label{scaling}
\end{equation}
The functions $f_{ij}$ depend on the invariant parton energy
$\sqrt{\hat s}$ and the gluino/squark masses. $\as$ is the QCD coupling
constant. For the sake of simplicity, the renormalization and
factorization scales are identified, $\mu_R = \mu_F = Q$. For the
non-zero scaling functions $f^{(0)}_{ij}$ to lowest order compact
expressions can be derived,
\begin{equation}
  f_{gg}^{(0)} = \frac{\pi\mg^2}{\hat{s}}\left\{\left[\frac{9}{4}
  +\frac{9\mg^2}{\hat{s}} -\frac{9\mg^4}{\hat{s}^2} \right]
  \log\left(\frac{1+\beta}{1-\beta} \right) -3\beta
  -\frac{51\beta\mg^2}{4\hat s} \right\}
\end{equation}
where $\beta = (1 -4\mg^2/\hat{s})^{1/2}$ denotes the gluino velocity in the
parton c.\,m.\,system. The scaling functions $f^{(1)}_{ij}$ and
$\bar{f}^{(1)}_{ij}$, which describe the next-to-leading order
corrections, are shown in Fig.~2a in terms of the variable $\eta =
\hat{s}/4\mg^2 -1$. The functions $f^{(1)}_{ij}$ are split into the
'virtual+soft' part (V+S) and the 'hard' part (H) in which the
infrared $\log^k\Delta~(k=1,2)$ singularities of the (V+S)
contribution are absorbed so that the limit $\Delta\to 0$ can be
carried out smoothly. The Sommerfeld rescattering contribution, due to
the exchange of Coulomb gluons between the slowly moving gluinos in
the final state, leads to
a singularity $\sim \pi\as/\beta$ near the threshold, which
compensates the phase space suppression. In addition, soft gluons give
rise to a logarithmic enhancement near threshold,
\begin{equation}
  f_{gg}^{(1)thr} = f^{(0)thr}_{gg} \left\{\frac{1}{16\beta} +
  \frac{3}{2\pi^2}\log^2(8\beta^2) -\frac{29}{4\pi^2}\log(8\beta^2) \right\}
\end{equation}
with
\begin{displaymath}
  f_{gg}^{(0)thr} = \frac{27 \pi\beta}{64}
\end{displaymath}
For large parton energies, the $\hat t$- and $\hat u$-channel
exchanges of gluons
generate an asymptotically constant cross section $\hat\sigma \sim
\as^3/\mg^2$. This is to be contrasted with the scaling behavior
$\hat\sigma_{LO} \sim \as^2/{\hat s}$ in lowest order. The asymptotic
values of the corresponding scaling functions are given by $
f_{gg}^{(1)} \to 1949/(800\pi)$ and $\bar{f}_{gg}^{(1)} \to
-177/(160\pi)$.

\subsection{Quark-antiquark initial states}

In addition to the annihilation via $\hat{s}$-channel gluon exchange,
also $\hat{t}$- and $\hat{u}$-channel squark exchanges contribute to
the process $q\bar{q}\to\gg$, Fig.~1b. To lowest order the cross
section is given by the function
\begin{eqnarray}
  f_{q\bar q}^{(0)} &=& \frac{\pi\mg^2}{\hat s} \left\{\beta\left[
  \frac{20}{27} + \frac{16 \mg^2}{9\hat s} -\frac{8\md^2}{3\hat s}
  +\frac{32 \md^4}{27(\md^4 +\ms^2\hat s)}\right] \right. \\
 & & \hphantom{\frac{\pi\mg^2}{\hat
     s}}\,+\left.\left[\frac{64\ms^2}{27\hat s} +\frac{8\md^4}{3\hat{s}^2}
 -\frac{16 \mg^2\md^2}{27 \hat s(\hat s -2\md^2)}\right]
 \log\left(\frac{1 -\beta -2\md^2/\hat{s}}{1 +\beta -2\md^2/\hat
   s}\right) \right\}
  \nonumber
\end{eqnarray}
where we have introduced the abbreviation $\md^2 = \mg^2 -\ms^2$. The
LO cross section is small for equal gluino and squark
masses due to the negative interference between the three diagrams. This
is also the case in next-to-leading order. The scaling functions
$f^{(1)}_{q\bar q}$ and
$\bar{f}^{(1)}_{q\bar q}$ associated with the next-to-leading order
corrections are displayed in Fig.~2b. At the threshold, the cross section
is non-zero due to the Sommerfeld enhancement,
\begin{equation}
    f_{q\bar q}^{(1)thr} = f_{q\bar q}^{(0)thr} \left\{\frac{3}{16\beta}
    +\frac{2}{3\pi^2}\log^2(8\beta^2)
    -\frac{41}{12\pi^2}\log(8\beta^2) \right\}
\end{equation}
with
\begin{displaymath}
    f_{q\bar q}^{(0)thr} = \frac{\pi\beta}{3}
  \left(\frac{\mg^2-\ms^2}{\mg^2+\ms^2}\right)^2
\end{displaymath}
Since any $\hat{t}$- or $\hat{u}$-channel gluon exchange diagrams are
absent, the scaling functions vanish for asymptotic energies.

\subsection{Quark-gluon initial states}
Gluino pairs can be produced in quark-gluon collisions only in higher
orders, $f^{(0)}_{qg}=0$, [cf.~Fig.~1d with the bottom gluon line
replaced by a quark line]. The corresponding scaling functions
$f^{(1)}_{qg}$ and $\bar{f}^{(1)}_{qg}$ are small, in the Tevatron
range, compared with the $q\bar q, gg$ scaling functions as shown in
Fig.~2c. The gluon exchange in the $\hat{t}$- and $\hat{u}$-channel
gives rise asymptotically to a non-vanishing cross section of order
$\as^3/\mg^2$ with $ f_{qg}^{(1)}\to 1949/(3600\pi)$ and
$\bar{f}_{qg}^{(1)} \to -59/(240\pi)$.

\section{Results}
The final results for the cross section $\ppb\to\gg X$ are presented
in Fig.~3 and Fig.~4. The cross sections of the subprocesses have been
convoluted with the parton densities in the GRV \cite{GRV},
CTEQ(2pM) \cite{CTEQ} and MRS(H) \cite{MRS} parametrizations. The detailed
analysis of these figures leads us to the following conclusions.

\noindent
{\bf (i)} It is obvious from Fig.~3a that the theoretical predictions for the
$\ppb\to\gg$ production process are greatly stabilized by taking into
account the next-to-leading order QCD corrections. While the
dependence on the renormalization/factorization scale $Q$ is quite
steep and monotonic in leading order, the $Q$ dependence is
significantly reduced in next-to-leading order for reasonable
variations of the scale. Close to $Q\sim \mg/2$ even a broad maximum
develops. The variation of the cross section for different NLO parton
parametrizations is very small.

\noindent
{\bf (ii)} In contrast to squark-pair production, the $K$-factors for
gluino pair production, $K = \sigma_{NLO}/\sigma_{LO}$ (with all
quantities in the numerator and denominator calculated consistently in
NLO and LO, respectively), depend rather strongly on the gluino and
squark masses, Fig.~3b, in particular in the range where squark and
gluino masses nearly coincide.

\noindent
{\bf (iii)} In Fig.~4 we illustrate the impact of the QCD corrections
on the experimental bound on the gluino mass. The scale
$\sqrt{\hat{s}}$ had been adopted in the experimental analyses, which
were based on LO cross sections. We therefore compare the LO cross
section, defined at this scale, with the NLO prediction evaluated at
the scale $Q=\mg$. As evident from the figure, this increases the
bound on the gluino mass by a shift between $10$ and $30$ GeV,
depending in detail on $\mg$ itself and on $\ms$. By comparing the NLO
prediction with the LO result in Fig.~4, it is clear that the bounds
extracted by the Tevatron experiments are conservative and that the
true bounds are likely to be higher at a level of $10$ to $30$ GeV.

\vspace{0.5cm}
{\bf Acknowledgment.} We thank S. Lammel for useful discussions on the
Tevatron squark and gluino mass limits.

\pagebreak

\pagebreak
\begin{figure}[h]
  \begin{center}
    \leavevmode
    \epsfig{file=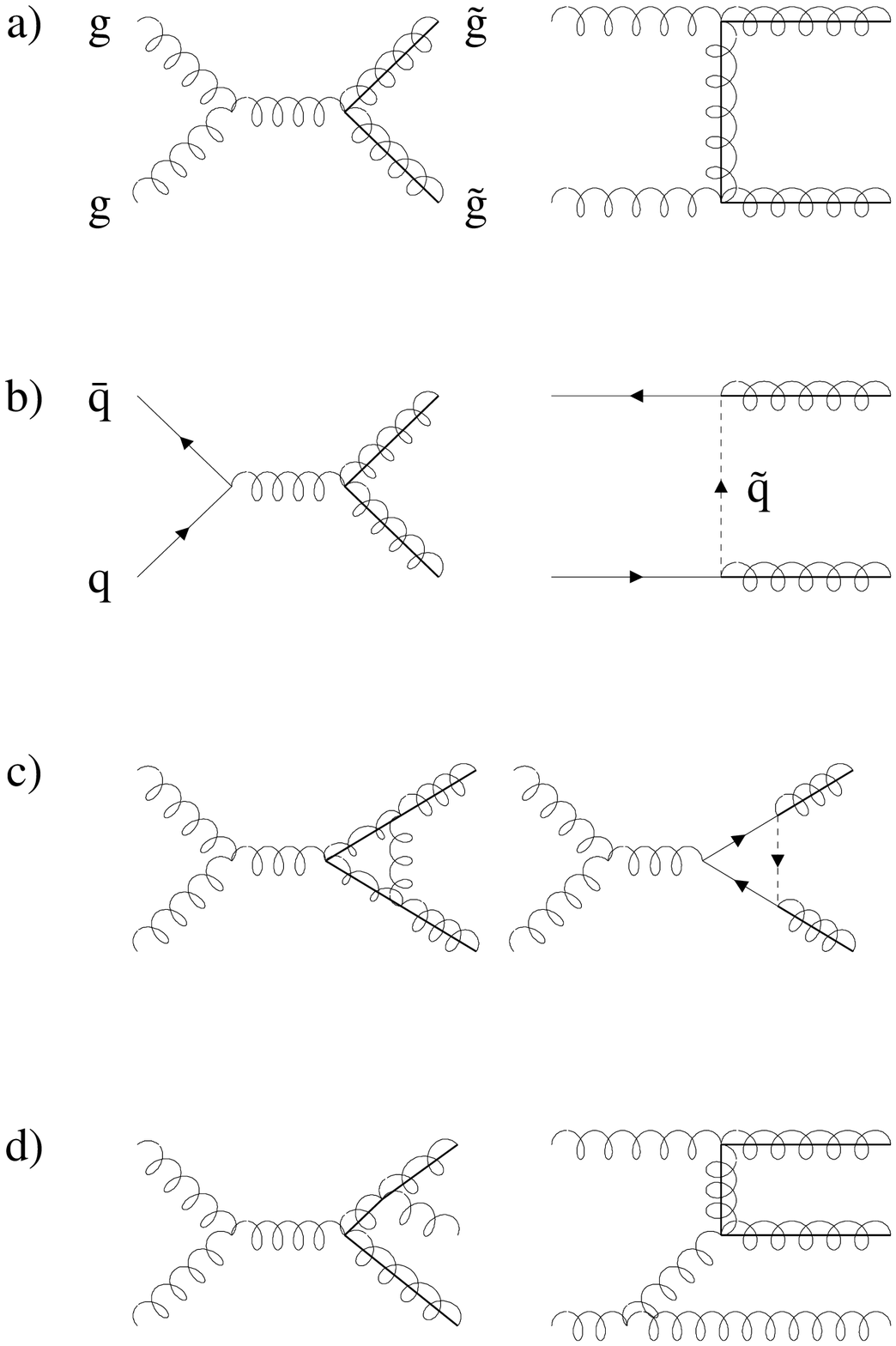,width=13cm}
  \end{center}
  \caption{Generic diagrams for gluino-pair production: a) Born-level
    $gg$ diagrams; b) Born-level $q\bar{q}$ diagrams; c) QCD and SUSY
    vertex corrections; d) Bremsstrahlung diagrams, and $qg$ parton
    processes if the bottom gluon line is replaced by a quark line.}
\end{figure}

\pagebreak
\begin{figure}[h]
  \begin{center}
    \leavevmode
    \epsfig{file=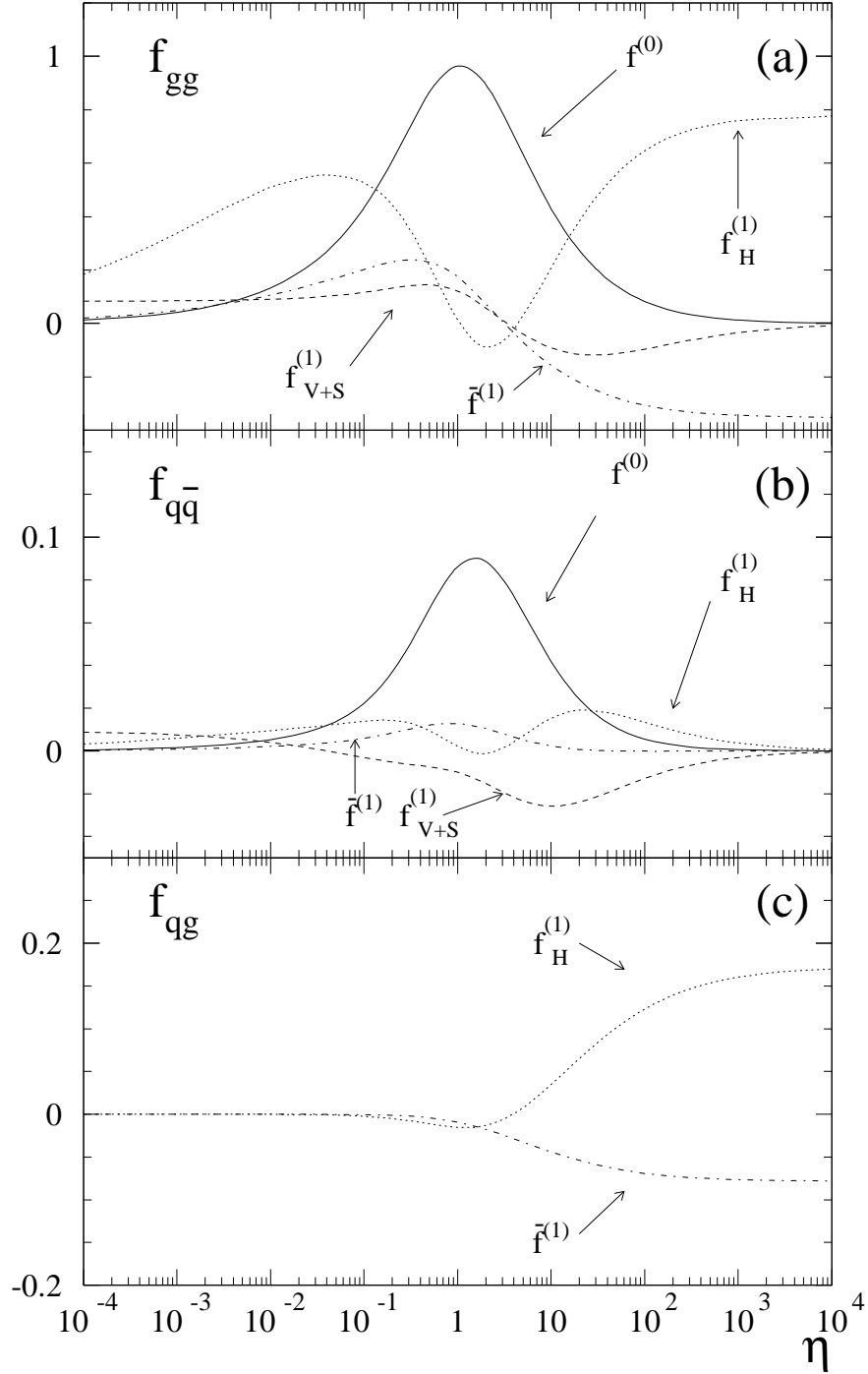,width=13cm}
  \end{center}
  \caption{The scaling functions for gluino-pair production in a) $gg$,
    b) $q\bar{q}$ and c) $qg$ collisisons. The notation follows
    eq.~(\protect{\ref{scaling}}) with $\eta = \hat{s}/4\mg^2 -1$; mass
    parameters: $\mg =250$ GeV and $\ms =200$ GeV.}
\end{figure}

\pagebreak
\begin{figure}[h]
  \begin{center}
    \leavevmode
    \epsfig{file=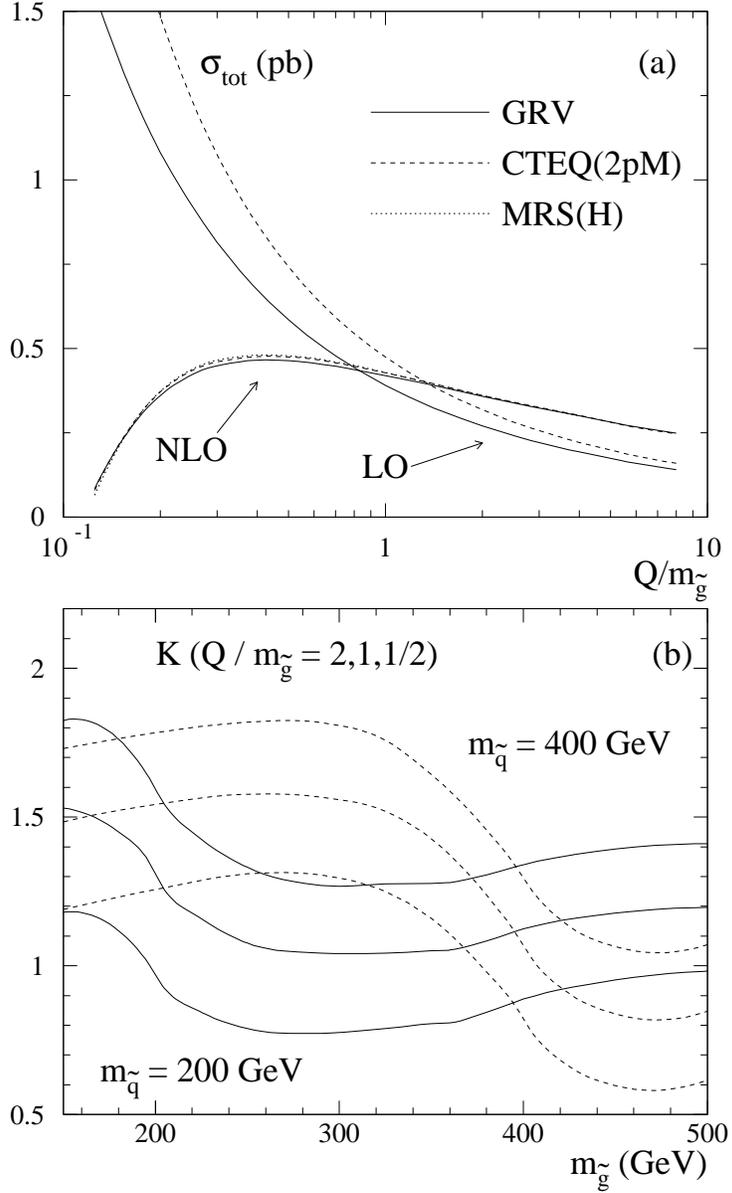,width=11cm}
  \end{center}
  \caption{Total cross section for gluino-pair production at the
    Tevatron energy $\protect{\sqrt{s}}=1.8$ TeV. a) Dependence on the
    scale $Q$ for the LO and NLO predictions, and sensitivity to
    different parton densities; mass parameters: $\mg =250$ GeV and
    $\ms =200$ GeV. b) K factors for the scales $Q/\mg=2,1,1/2$,
    corresponding to upper, middle and lower curves, respectively; GRV
    parton densities.}
\end{figure}

\pagebreak
\begin{figure}[h]
  \begin{center}
    \leavevmode
    \epsfig{file=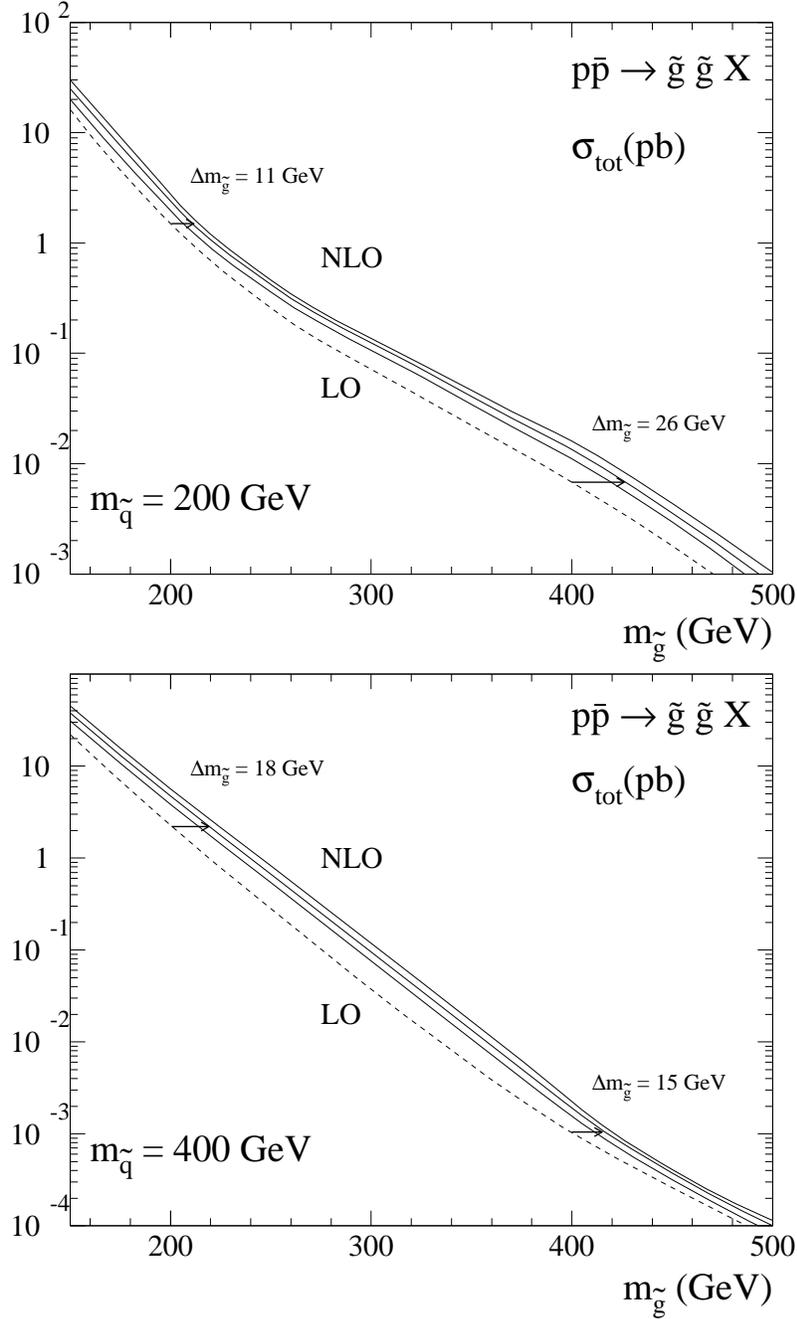,width=12cm}
  \end{center}
  \caption{Total cross section for gluino-pair production at the
    Tevatron energy $\protect{\sqrt{s}}=1.8$ TeV. Dependence of the
    cross section on the gluino mass; LO with EHLQ parton densities and
    $Q =\protect{\sqrt{\hat{s}}}$ and NLO with GRV parton densities at
    the scales $Q/\mg =2,1,1/2$ corresponding to lower, middle and
    upper curve, respectively.}
\end{figure}

\end{document}